%% file: context.tex
\documentclass[letterpaper,twocolumn,10pt]{article}
\usepackage{usenix}

\usepackage{array, graphics, graphicx, color, xspace, url, multirow, rotating, epsfig, amsmath, amsfonts, subfig}
\usepackage{wrapfig}
\usepackage{outlines} 
\usepackage{tikz}
\usepackage{epstopdf,subfig}
\usepackage{multicol}
\usepackage[format=plain,labelfont=bf,font=small]{caption}
\usepackage{ragged2e,flushend,enumitem,listings}
\usepackage[english]{babel}
\usepackage{blindtext}
\usepackage{acronym} 
\usepackage{breakurl}
\usepackage{hyperref}
\usepackage{cleveref}
\usepackage[frozencache,cachedir=.]{minted}
\usepackage{appendix}
\usepackage{filecontents}
\usepackage[compact,small]{titlesec}
\usepackage{pifont}%
\usepackage{algorithm}
\usepackage[noend]{algpseudocode}
\algrenewcommand\alglinenumber[1]{\large #1:}

\newcommand{\squishlist}{
  \begin{list}{$\bullet$}{
    \setlength{\itemsep}{0pt}       \setlength{\parsep}{0pt}
    \setlength{\topsep}{0pt}        \setlength{\partopsep}{0pt}
    \setlength{\leftmargin}{1em}    \setlength{\labelwidth}{1em}
    \setlength{\labelsep}{0.5em} } }
\newcommand{\squishend}{
  \end{list} }
\newcommand{\squishlistend}{
  \end{list} }
  
\crefformat{section}{\S#2#1#3}
\crefformat{subsection}{\S#2#1#3}
\crefformat{subsubsection}{\S#2#1#3}

\setminted{fontsize=\footnotesize, xleftmargin=0.22in, fontsize=\footnotesize, linenos}
\makeatletter
\newcommand{\currentfontsize}{\fontsize{\f@size}{\f@baselineskip}\selectfont}
\makeatother
\BeforeBeginEnvironment{minted}{\medskip}

\newcommand{\bi}{\begin{itemize}}
\newcommand{\ei}{\end{itemize}}

\newcommand{\etc}{{\it etc.}\xspace}
\newcommand{\ie}{\emph{i.e.,}\xspace}
\newcommand{\eg}{\emph{e.g.,}\xspace}

\newcommand{\vs}{{\it vs.}\xspace}

\newcommand\paragraphb[1]{\noindent{\bf{#1}}}

\newcommand\pb[1]{\paragraphb{#1}}

\newcommand{\figspace}{\vspace{-10pt}}
\newcommand{\minttopspace}{\vspace{-0.07in}}

\newcommand{\jitda}{{JIT-DA}\xspace}
\newcommand{\jit}{{JIT}\xspace}
\newcommand{\arch}{{Jut}\xspace} %

\newcommand{\cdrs}{{CDRs}\xspace}

\newcommand{\join}{{\texttt{join}}\xspace}
\newcommand{\leave}{{\texttt{leave}}\xspace}

\newcommand{\Au}{{$A_{user}$}\xspace}
\newcommand{\Ab}{{$A_{bldg}$}\xspace}
\newcommand{\Cu}{{$C_{user}$}\xspace}
\newcommand{\Cb}{{$C_{bldg}$}\xspace}

\definecolor{codegray}{rgb}{0.5,0.5,0.5}
\lstdefinestyle{bashStyle}{
    commentstyle=\color{blue},
    numberstyle=\tiny\color{codegray},
    basicstyle=\ttfamily\scriptsize,
    breakatwhitespace=false,         
    breaklines=true,                 
    captionpos=b,                    
    keepspaces=true,                 
    numbers=left,                    
    numbersep=5pt,                  
    showspaces=false,                
    showstringspaces=false,
    showtabs=false,                  
    tabsize=2,
    keywordstyle=\bfseries,
}

\date{}
\graphicspath{{./dia/}}

\newcommand{\eat}[1]{}

\begin{document}
\title{\arch: A Framework for Just-in-Time Data Access}

\author{Silvery Fu$^1$, Siyuan Dong$^1$, Jamsheed Mistri$^1$, Amy Ousterhout$^2$, Steve McCanne$^3$, Sylvia Ratnasamy$^1$
\\
\textrm{$^1$UC Berkeley \hspace{0.05 in} $^2$UCSD \hspace{0.05 in} $^3$Brim Data}
}
\maketitle

\begin{abstract}
With the proliferation of sensor and personal devices, our physical spaces are now awash in potential data sources. In principle this data could serve a wide range of applications and services. However, leveraging these data sources is challenging with today’s systems because they are not typically designed to consume data \emph{opportunistically}, from a new device that happens to arrive in the vicinity. 

In this paper, we present the design and implementation of \arch, a system designed for ``Just-in-Time’’ data access — in which an application is able to discover and consume data from \emph{any} available source, even ones not known at development or installation time. \arch combines two novel design choices:  \emph{modularizing} data processing systems to better reflect the physical world, and a new form of application-data \emph{integration} that equips data processing pipelines with the information they need to process new and evolving data formats and schemas. We show that these choices greatly simplify the development and use of smart-space and IoT applications. For a representative set of devices and application scenarios, we show that \arch can implement use-cases not easily supported today, or can do so with 3.2-14.8$\times$ less development effort and 3-12$\times$ lower query complexity than current systems.

\end{abstract}

\input{intro}

\input{s2}
\input{s3}
\input{s4} 
\input{s5}

\input{related}

\input{concl}
\onecolumn \begin{multicols}{2}
\bibliographystyle{abbrv} 
\begin{small}
\bibliography{context}
\end{small}
\end{multicols}

\twocolumn
\input{supplement}

\end{document}

%% file: intro.tex
\section{Introduction}
\label{sec:intro}
From phones to health trackers, smart appliances, and self-driving cars, we are witnessing an explosive growth in connected devices. These devices stand to transform the world of data since every connected device is now a potential data source. And while early Big Data was driven by the growth in online content (\eg web pages, video), today, data is increasingly generated by devices in the \emph{physical} world. 

A growing number of applications in domains ranging from retail and heath, to smart spaces and environmental monitoring would like to leverage these data sources.
An important class of such applications --- and our primary focus in this paper --- relates to smart spaces. These apps must accurately capture our physical spaces and hence leveraging data from users and devices that inhabit the space is invaluable.
E.g., consider a smart campus app that today relies on data from sensors deployed by the building owners to optimize their HVAC systems for current occupancy and air quality. In principle, this app’s dataset could be augmented by data collected from the devices carried by the space’s occupants. 
I.e., a user entering a building may contribute sensor readings from her phone to the building’s app. This data could improve the app's accuracy (offering data from a different vantage point) and cost (requiring fewer sensors). %

For the most part, today’s apps are limited to data from sources that are predefined and manually onboarded rather than (say) collected from a user that happened to walk in the door. Our goal is to enable apps to consume data in a manner that is more opportunistic than organized: apps should dynamically discover potential sources, ingest their data, and incorporate this data into the data-driven insights they expose. We call this ``\emph{Just-in-Time}'' data access (\jitda), reflecting the fact that we want to leverage data sources discovered at runtime \vs  development or installation time. 

Can we achieve \jitda with today's systems? The current landscape of smart-space and IoT app development is, unfortunately, rather messy. These apps are often closed and vertical silos, with a single vendor that develops an app to manage their own devices~\cite{smartthings, lifx-light,nest} - as such, they don’t address the problem of \jit data sources. An alternative is to use one of the open frameworks for IoT app development - \eg AWS IoT~\cite{aws-iot} and Home Assistant~\cite{homeassistant} in the commercial arena or research prototypes such as BOSS~\cite{nsdi13-boss} and dSpace~\cite{dspace}. These are \emph{application} frameworks - they provide services~\cite{aws-iot,aws-iot-things-graph} or code~\cite{lifxlan,apple-homekit,matter} that aid in onboarding devices and expressing the application logic for automating these devices: \eg defining IFTTT rules~\cite{ifttt} for automation (\eg if <condition> then dim lights), defining identifier and tag namespaces (\eg to represent rooms, buildings), associating tags with devices (\eg tagging a lamp with the id of the room it is in), grouping devices (\eg lamps in room Foo), and so forth. The focus of these app frameworks is on device automation and actuation logic and, as such, their support for data storage and analytics is thin. The typical assumption is that data streams from devices will be loaded into a standalone database/datalake. Leveraging insights from this data is typically out of scope for these app frameworks and hence it is left to the app developer to bring in a separate data processing system\footnote{By a data processing system, we mean the various components to do with data ingestion (parsing, cleaning, normalizing input data), storage (\eg row- or column-based),  organization (\eg schema registries, metadata tables, indexes, repositories), query engines, and so forth.}  and integrate it with their app, a process that developers typically address in a bespoke manner. Relevant to our focus:  in both app and data processing frameworks today, the operator typically onboards new devices or data sources manually and, as such, they are not designed for \jit data. 

The \arch framework we present in this paper aims to enable the use of \jit data sources in smartspace/IoT apps. As we’ll discuss, the manner in which \arch achieves this has the fortuitous benefit of simplifying the above app development process, even for regular (\ie non-\jit) data sources. In a nutshell, \arch achieves this by: (i) providing a richer and more systematic approach to integrating app and data processing layers and, (ii) making it easier for app developers to efficiently identify and query data that is relevant to what we term an application-level \emph{context}, such as a room or building. In the remainder of this section, we elaborate on both the gaps in existing systems when it comes to \jit data sources and how \arch addresses these gaps.

\subsection{Challenges in realizing \jitda}
\label{subsec:challenges}
In this section, we argue that enabling \jitda must start by examining the data processing architecture that typically underlies data-driven apps.
Realizing \jitda with existing data processing systems is hard for three key reasons.

\paragraph{(1) Unplanned data sources.} \jit data sources are often unplanned in the sense that they are not known at app development or installation time. As a result, the app developer is unaware of both what \emph{form} the input data will take in terms of data formats and schemas, and \emph{when}, if at all, the data source will appear.
Instead, as users in (for example) our smart campus enter the building, their data streams must be integrated with the building’s ingestion pipeline, appear in the query results, \etc

As a result, the data processing pipeline must be prepared to handle data that is \emph{heterogeneous} (spanning different representations and semantics) and must be \emph{dynamic} (evolving at runtime in an automated manner to handle new sources that appear/disappear). 
As we discuss in \S\ref{sec:s2}, recent efforts in the database community address the former but not the latter~\cite{zed-cidr,proteus,sparser,redbook_integration}. Instead, in today’s systems, the data processing pipeline is typically static and evolution requires manual intervention. Our work addresses how we can construct pipelines that adapt dynamically and automatically to app-level events, while incorporating recent techniques for heterogeneous data models within these pipelines. 

\paragraph{(2) Decentralized ownership and policies.} 
 In the scenarios we consider, an app’s data sources might be owned and operated independently from the app provider. This is true not just of end users and their devices but also across smart-space operators - \eg even within a single campus, the operator of a CS building might have very different operational requirements from that of the life sciences building. Thus in such app environments, supporting the policies and autonomy of the various stakeholders becomes key. A user might share data with one building app but not another; one building operator might have restrictions on whether their data can be stored in the cloud while another may not; \etc Today, an app's input streams are typically dumped into a monolithic database or data lake, making it difficult to impose fine-grained policies - \eg per building, department, or user.

\paragraph{(3) Data sources are application unaware.} 
The apps we consider typically have abstractions that reflect their application context -- \eg room, building -- and the entire purpose of these apps is to support operations over these abstractions: querying "the room", configuring "the building", and so forth. However, because \jitda sources are independent of a space or app, these abstractions are typically not represented in their data schemas: \eg the data schema that a phone vendor uses to represent the temperature readings on the phone are unlikely to include a field that represents the building the phone is currently in. 
This omission complicates querying "the room" because a necessary first step in querying a context is knowing which data sources are \emph{associated} with that context: \eg a query for the average room temperature must first know which phones were present in the room and when, so that it only uses data from the appropriate phones at the appropriate times.

Unfortunately, as mentioned above, the \emph{association} between an app and a \jit data source is potentially complex: opportunistic, dynamically changing, and subject to distributed control. And yet, \emph{while this association lasts}, we want to incorporate the source's data into the application logic. This implies that someone must track the association between the app and its data sources. 
Today's app frameworks offer no systematic support for tracking these associations nor for translating
from operations on high-level app abstractions to a heterogeneous and evolving set of per-source data streams.
Instead app developers must do so with bespoke and ad-hoc techniques: \eg defining metadata stores that track when and which sources were in the room, rewriting queries, dynamically augmenting the input data stream. %
 As we illustrate in \S\ref{sec:eval}, this adds significant complexity to the development of smart-space apps.%

\subsection{\arch: An architecture for \jitda}
In this paper we present \arch, a new framework for building \jitda apps that addresses the above challenges. 
Our approach starts with the observation that the abstractions (\eg devices, rooms, buildings) and information (\eg when a device is present in a room) needed to address some of our challenges can be found in the \emph{application} layer. The problem is that these abstractions/information are absent or not well represented in the underlying data processing systems. Moreover, there is no way to systematically and automatically reflect this app-level information in the underlying data processing layer; \eg when the app detects a new device (described below), this should automatically trigger the app's data processing pipeline to consider ingesting data from the device's APIs. 

Our framework thus adapts existing data architectures as follows.
First, we modularize the data processing layer to reflect the natural modularity found in smart-space apps. Thus, instead of one monolithic data processing pipeline, in \arch, 
every entity -- whether a device data source, a room, or a building -- has an independent data processing pipeline that implements an ingest dataflow (\ie  processing input data relevant to that entity), storage, and an egress dataflow (\ie for exporting data from its store). We refer to an entity modeled in this way as a \emph{context}.

Next, we extend the data processing architecture to expose a new \texttt{join(C1, C2)} interface (and corresponding \texttt{leave()}) via which an application can inform its underlying data processing system that a context C1 is available to serve as a data source for a context C2. The implementation of \join connects C1’s egress pipeline to C2’s ingestion pipeline. Importantly, \texttt{join} does so in a manner that enforces their respective policies (\eg that $C1$ is from a trusted vendor) and reconciles any mismatch in their heterogeneous data representations (\eg converting from $C1$'s data format to that required by $C2$'s storage system). Thus a key contribution in \arch is a novel approach that, at runtime, compiles high-level policies into low-level functions for processing newly discovered data source. 
In effect, \join allows applications to "plumb" contexts, to reflect the (dynamically evolving) relationship between \jit data sources and their physical space or context.

Taken together, the above changes address the challenges from \S\ref{subsec:challenges}. The modularity of contexts means we can develop and operate context code independently, supporting decentralized ownership and control. In addition, this modularity allows us to efficiently implement “by context” queries, since a context’s data is curated in its datastore (\vs requiring additional steps to first identify data relevant to a context). 
Finally, the \join interface enables data processing pipelines that adapt to reflect the complex interactions between data sources and consumers in the physical world.

\arch thus provides general yet powerful abstractions that simplify building (smart-space) apps. Specifically, an app developer interacts with the \arch framework to create contexts, compose, and query them. Because contexts are an explicit yet modular and composable abstraction, developers can: (i) treat contexts themselves as data sources that can be queried, logged, configured, \etc, (ii) can compose contexts into higher level ones (\eg rooms as data sources in a building context); (iii) reuse contexts to accelerate development, and so forth.

\arch's changes do not impact the internal techniques of current data processing systems: \eg their storage formats, query planners, optimizers.
Instead, the novelty that \arch brings is a shift in the \emph{system} architecture of data processing systems: introducing a new modularity (based on contexts) and a new, more systematic and structured, integration between data and app layers (based on joining contexts). 

In this paper, we present the design and implementation of the \arch framework that implements the above design approach. We evaluate \arch by using it to implement a range of smart-space scenarios using real physical devices and spaces.
Our results show that \arch can implement use-cases that are not easily supported today, or can do so with 3.2-14.8$\times$ less development effort and 3-12$\times$ lower query complexity than current commercial~\cite{aws-iot} and research~\cite{dspace} systems.

The remainder of this paper is organized as follows: we elaborate on our goals and assumptions in \S\ref{sec:s2}, then present \arch's design and implementation in \S\ref{sec:s3} and \S\ref{sec:impl} respectively. We evaluate \arch in \S\ref{sec:eval} and then conclude.

%% file: s2.tex
\section{Goals and Assumptions}
\label{sec:s2}
Our goal is to make it easier for app developers to leverage the wealth of data sources available to them. We focus on smart-space apps as they are an important emerging class of applications,\footnote{Reports project the smart space market will cross \$100B by 2030, fueled by the deployment of mobile and IoT devices and the adoption of AI techniques to consume/control these devices.} %
but also because the number of connected devices in such spaces has grown tremendously and yet today's apps leverage data from only a small fraction of them. 

\paragraph{Operator and user incentives.}
Our work is predicated on two assumptions regarding incentives. The first is that providers of smart-space apps are interested in consuming data from 3rd-party data sources; \ie in \jitda. We believe this assumption is reasonable for a few different reasons. First, the additional data can improve the quality of the insights these apps provide. For example, by using data from tenants' BYOD devices~\cite{byod} such as smartphones and laptops, a smart building app such as Comfy~\cite{comfy} can estimate building occupancy, monitor ambient noise levels, and assess network quality in real-time across all occupied areas. Second, leveraging third-party data sources means that operators can deploy fewer data sensors (cameras, air quality monitors, motion sensors, and so on) reducing both their capital expenditure and the operational costs of maintaining the same. For example, tracking the number of people throughout a commercial building (typically with 100s of rooms) can be cost-prohibitive when using fixed, dedicated sensors like the Density radar~\cite{density}. Each radar, designed to cover around 1,325 sq. ft. for a room or space, has an upfront installation fee of \$895, accompanied by an annual maintenance and data access charge of \$795~\cite{density-product}. The total annual operational costs can easily exceed \$100,000, not to mention for campuses or complexes spanning 1,000s of acres with many buildings. In contrast, leveraging data from tenants' devices can augment these dedicated sensors or even replace them in areas, achieving similar objectives at a fraction of the cost.
Finally, such data can enable new differentiating features that are otherwise not possible. For example, by leveraging data about tenants' exercise habits  (\eg Fitbit steps taken~\cite{fitbit}), a building app can suggest appropriate walking routes and highlight available amenities within the building.

\paragraph{User incentives.}
Likewise, we assume that users/sources may opt to share data with some smart-space app providers. We envision multiple possibilities why they might do so. One is because it leads to reciprocal benefits - \ie sharing my data improves my app experience. For example, the smart building app could tag building areas based on tenant activity data, while recommending them quiet rooms for deep work. Another is that the user might have no option but to do so given the terms of their app/device. This is often the case, for example, with corporate apps and devices where employees may be required to use specific apps that track usage, location, or other data as part of the organization's IT security and compliance measures. Yet another possibility is that the user benefits financially from sharing their data. For example, employees might be offered discounts on building amenities, in exchange for sharing certain data. Finally, in some situations, users might volunteer their data for purely altruistic motives. Consider a scenario where building occupants voluntarily share data to monitor environmental conditions, aiming for a communal goal of reducing the building's carbon footprint or promoting sustainability.%

\paragraph{Service discovery.} 
Our work also makes two assumptions on the technical front. 
The first is that service discovery - \ie discovering apps and their APIs - is largely a solved problem. Consider our canonical scenario in which a user, running an app A1 on her device, walks into a building that is associated with a smart-space app A2. Our ultimate goal is to incorporate A1's data stream into A2's analytics pipeline and/or vice versa. But first A1 and A2 must discover each other based on location and exchange information about their APIs, data schemas, and so forth. 
This discovery can be bootstrapped through one of many well-known techniques: (i) the user scans a QR code in the building~\cite{qr}, (ii) a captive portal on attaching to the building’s WiFi AP~\cite{wifi-login}, (iii) a well-known naming scheme (\eg building.cs.univ.edu), and so forth. 
 Once this initial discovery is complete, A1 and A2 can exchange higher-level information (\eg API descriptions, credentials, schemas) in a straightforward manner via techniques such as: (i) a cloud-based directory service(s) that A1 and A2 register with, (ii) a peer-to-peer protocol in which A1 and A2 directly exchange relevant information, and so forth. API discovery of this form is common even today in systems such as GCP service directory~\cite{service-directory}, RapidAPI~\cite{rapidapi}, BLE neighbor discovery~\cite{ble-modeling}, and AirDrop~\cite{airdrop}.

\paragraph{Heterogeneous data.} 
Our second technical assumption is that data analytics over heterogeneous data is, or will soon be, a solved problem. Processing heterogeneous data, that spans different schemas and formats, has been a long-standing challenge in databases~\cite{redbook_intro,doan2012principles}. Heterogeneity raises two high-level problems. The first is that querying across heterogeneous data must reconcile the \emph{syntactic and semantic} gaps across different entries in the dataset~\cite{redbook_integration}. The former occurs when different formats are used to represent identical information (\eg string vs. int types; or parquet vs. json) while the latter refers to fields that define related but not identical information. In some cases, we can reconcile a semantic gap (\eg converting a field in celsius to fahrenheit) while others prove harder (\eg a sensor that defines its temperature field as the average temperature \vs one that defines it as the max). Reconciling such gaps is a fundamentally difficult problem. Yet both research and deployed systems have adopted a range of techniques that mitigate (if not ``solve'') the issue: formal or de-facto naming standards~\cite{li2013naming}, manually coded or auto-generated format converters~\cite{doan2012principles}, AI-driven translation functions~\cite{data-int}, ontologies and schema registries~\cite{schema-registry,buildsys16-brick}, \etc. As we discuss in \S\ref{subsec:join}, our work assumes such techniques exist and shows how to use them in a \jitda environment.

\begin{outline}

\end{outline}

%% file: s3.tex
\section{Design}
\label{sec:s3}
In this section, we first present the design of our \arch framework (\S\ref{subsec:context} and \S\ref{subsec:join}) and then step through an end-to-end example of how a \jitda app  operates, highlighting how our approach differs from current systems (\S\ref{subsec:e2e}).\footnote{In \S\ref{sec:eval}, we compare building an app using \arch{} \vs doing so with existing solutions: AWS IoT over AWS Timestream and dSpace over Postgres.}

\arch is a data processing architecture that enables app developers to opportunistically leverage data sources present in a physical environment. Such developers may implement their application logic using an app framework (\eg AWS-IoT, dSpace, SmartThings) but will leverage \arch as the data storage and processing system underlying these applications.%

There are two important pieces to the \arch architecture: (1) the abstraction of a \emph{context} and, (2) the \join (correspondingly \leave) API used to compose contexts in reaction to app-level events. We discuss each in turn below.

\begin{table}[]
\footnotesize
\centering
\begin{tabular}{|c|c|c|c|}
\hline
\textbf{Name} & \textbf{Notation} & \textbf{API} & \textbf{Description} \\ \hline
\multirow{4}{*}{Metadata} & \multirow{4}{*}{-} & .kind & Kind of the context \\ \cline{3-4} 
 &  & .name & Name of the context \\ \cline{3-4} 
 &  & .intent & Contexts to ingest data from \\ \cline{3-4} 
 &  & .role & Role for access control \\ \hline
\multirow{2}{*}{\begin{tabular}[c]{@{}c@{}}Data\\ Store\end{tabular}} & \multirow{2}{*}{$C.store$} & load() & Load data to the store \\ \cline{3-4} 
 &  & query() & Query the store \\ \hline
\multirow{2}{*}{Egress} & \multirow{2}{*}{$C.egress$} & .id & Egress identifier \\ \cline{3-4} 
 &  & .view & \begin{tabular}[c]{@{}c@{}}Process and cache \\ derived data from the store\end{tabular} \\ \hline
\multirow{4}{*}{Ingress} & \multirow{4}{*}{$C.ingress$} & .id & Ingress identifier \\ \cline{3-4} 
 &  & .source & Data source of the ingress \\ \cline{3-4} 
 &  & .rule & Schema match-action rules \\ \cline{3-4} 
 &  & .flow & \begin{tabular}[c]{@{}c@{}}Operators that process \\ data from the source\end{tabular} \\ \hline
\end{tabular}
\caption{Main components and their APIs in a context.}
\label{tab:abs}

\figspace
\end{table}

\subsection{The \emph{context} abstraction}
\label{subsec:context}
Contexts impose modularity in the data processing layer, reflecting the modularity commonly found in the app layer.

\paragraph{Context metadata.}  A context $C$ has three important pieces of associated metadata : \emph{names}, \emph{kinds}, and sourcing \emph{intents}.  

The \emph{name} field acts as a unique identifier for $C$ while its \emph{kind} is used to identify the broader class of contexts to which $C$ belongs. For example, all Apple smartphones might share the same \emph{kind="iphone"} while a user might use the \emph{name} field to identify their specific instance of an iphone. In this sense, a context \emph{kind} resembles a ``class'' in programming languages, allowing for the creation and customization of instances based on that kind. The rationale for including both names and kinds is that while there will be an enormous number of devices/contexts, there will be far fewer \emph{kinds} of devices/contexts that will be of interest to a specific context. Developers can write data processing pipelines  for these kinds, and this approach encourages a natural consolidation towards a few popular/standard kinds (\eg apple.com/v1/iphone) while still giving developers the flexibility to support more niche kinds (vendor\_x/v1/phone). 
Moreover, \emph{kinds} allow developers to indicate their ability to ingest data from classes of sources rather than specific ones; this is desirable since specific contexts might not be known at the development time.

Finally sourcing \emph{intents} list the other contexts that $C$ is prepared to ingest data from. Intents are specified in terms of the \emph{names} and \emph{kinds} (including wildcards) of other contexts thus allowing a developer to limit ingestion to a specific other context (\eg name="building.cs.foo-univ.edu") or to a potentially large set of them (\eg kind="apple.iphone"; kind="\*"). Context metadata are summarized in Table~\ref{tab:abs}.

\paragraph{Components of a context.} A context $C$ is composed of three components: a data store (\emph{C.store}), an ingress (\emph{C.ingress}), and an egress pipeline (\emph{C.egress}). 

\noindent \underline{\emph{C.store}} is a regular data store (whether relational database, data lake, or lakehouse\footnote{For our implementation, we choose a particular existing datastore that we use with no modification (\S\ref{sec:impl}).}) that is used to store data relevant to $C$. It thus acts as a convenient and efficient repository over which to execute queries pertaining to $C$.

\noindent \underline{\emph{C.egress}} 
A context $C$ can have one or more egresses, each of which defines a schema representing data that $C$ exports. Thus an egress exposes data records in $C.store$ to data consumers such as other contexts, apps, and users. $C.egress$ can also be associated with a dataflow pipeline:  at runtime, $C.egress$ reads from $C.store$, processes records as per the dataflow operators, and then caches the results persistently. We refer to this resultant data as the \emph{egress view}, borrowing the notion of views from databases~\cite{noria,risingwave,postgres}. Like any database view, $C.egress$ can be queried and our implementation supports both one-shot queries (that return a set of records) and continuous queries (that return a stream of records). Table~\ref{tab:abs} summarizes the relevant fields in $C.egress$. 

\noindent \underline{\emph{C.ingress}} 
An ingress reads data from a source, processes this data, and stores it in $C.store$.  A context $C$ can have multiple ingresses, each with a unique identifier.
The sources associated with a particular ingress are determined during the \texttt{join} process which we describe shortly in \S\ref{subsec:join}. 
Processing is implemented as a dataflow which contains operators (\eg sort, join) and functions (\eg sum, filter) that process a sequence of input data records and generate a sequence of output records. The resultant \emph{derived} data is written to $C.store$. Again, deciding which dataflow operators and functions should be applied to a given source is done during the \texttt{join} process. 

An important aspect of an ingress is a "match:action" table that expresses high-level rules/policies for how the ingress will handle heterogeneous data schemas.  
An entry in the table specifies the conditions for matching a schema and the corresponding actions to be taken. For example:
\[
\begin{tabular}{ll}
\textbf{Match} & \textbf{Action} \\
\text{has <watt: string>} & \text{extract} \\
\text{has <power: string>} & \text{rename watt:=power} \\
* & \text{reject} \\
\end{tabular}
\]

These rules are used to update the ingress's dataflow at runtime, in response to join events as described below.
Our current implementation allows matching on: schema name, combinations of schema fields (\texttt{all}, \texttt{any}),  and wildcards. Actions include developer-provided functions such as \texttt{drop} (the record will not be processed and stored), \texttt{log} (routes the record to a separate log store), \texttt{trim} (deletes fields), \texttt{rename} (for field names), custom field conversions (\eg fahrenheit to celsius), and so forth. Our appendices include additional detail.
More generally, we expect three categories of actions: (i) accepting the data as is, (ii) rejecting all data if unexpected fields are encountered, or (iii) transforming the data to adhere to a known schema, such as by filtering out unknown fields or converting them to a predefined schema.

Both match and actions are extensible allowing developers to tailor the system to their application domain.

\begin{outline}
\end{outline}

\subsection{The \texttt{join}/\texttt{leave} interface}
As mentioned earlier, contexts can be composed which allows data to flow from the egress of one context to the ingress of another: \eg the egress of a user device might be composed with the ingress of a room context and, simultaneously, the egress of the room context might be composed with the ingress of the building context. 

The \texttt{join} (\texttt{leave}) API is how such composition (decomposition) is realized. As described in \S\ref{sec:s2}, we assume that two applications can discover one another through some service discovery mechanism. 

For illustration, we'll consider two applications $A_{user}$ and $A_{bldg}$ that have discovered each other: \eg in our canonical scenario, $A_{user}$ might be an app on a user's phone that exports sensor readings and other information from the phone, while $A_{bldg}$ is the building app for the building that the user just entered. Let's say that $C_{user}$ and $C_{bldg}$ are the contexts associated with $A_user$ and $A_{bldg}$ respectively. As part of the discovery process, we'll assume that $A_{user}$ and $A_{bldg}$ exchange information about their respective contexts and focus on the actions taken at $A_{bldg}$; processing at $A_{user}$ to incorporate $C_{bldg}$ as a data source in $C_{user}$ follows along the same lines. We assume that $A_{bldg}$ will first perform any application-level security and policy checks deemed necessary and, assuming these are satisfied, $A_{bldg}$ then invokes its underlying \arch layer's \join interface thus letting $C_{bldg}$ know that $C_{user}$ is available as a potential data source. %

To implement \join, the \arch runtime first checks whether $C_{user}$ is a potential data source for $C_{bldg}$ by comparing $C_{user}$'s \emph{name} and \emph{kind} fields against $C_{bldg}$'s sourcing \emph{intents}. In case of a match, it then checks each of $C_{user}$’s egress schemas against each of $C_{bldg}$’s ingress match:action tables. If the schema for one of $C_{user}$’s egresses, denoted $C_{user}.egress.idx$ matches a rule in the match:action table for $C_{bldg}$’s ingress $C_{bldg}.ingress.idy$ then the corresponding actions are compiled into dataflow operators and the generated dataflow is prepended to the existing dataflow for $C_{bldg}.ingress.idy$ and $C_{user}.egresss.idx$ is added as a data source for $C_{bldg}$. For instance, given a schema \texttt{<measurement:watt:string>}, the first rule from the table above is matched, and a dataflow is added to $ingress$'s dataflow that extracts the field \texttt{watt} from data records matching the schema. 

In summary, using the above \join (and corresponding \leave) interface we can ``plumb" data sources to contexts, and contexts to other contexts. Importantly, the use of these interfaces is not limited to \jit sources; rather developers can use them to plumb any data source to the context - whether a \jit phone, a static building sensor, or another known context. 
Thus, using \join/\leave gives developers a uniform approach to setting up data processing pipelines that capture the potentially complex and evolving association between data sources and context apps.

Now that we've explained how contexts and \join work, we briefly discuss the role of naming. 
\arch does not require that data sources adopt a single naming scheme but it certainly works better if data sources consolidate around a smaller number of approaches. 
Specifically, in deciding whether and how to ingest data from a source, \arch uses a context's \emph{name} and \emph{kind} fields, together with the data schemas for its egresses. Any of these are optional and developers can always choose whether to accept or reject data with an unknown name, type, or schema. Likewise, the developer of a context can always choose whether to embrace an existing naming scheme or define their own. 
In practice, we expect that a small number of naming schemes will emerge as defacto standards with widely available libraries to convert between them. This is, in fact, how the world of data works even today with defacto formats such as JSON, Parquet, \etc; protocols such as Matter for unified device communication~\cite{matter}; Project Haystack~\cite{haystack} and Brick for building ontologies~\cite{buildsys16-brick}, and so forth.
\arch's contribution is in taking these techniques (formats, schema converters, \etc) from  the database literature and showing how to automatically and systematically insert them into the processing pipeline as new \jit sources appear/disappear.

More generally, bridging the syntactic/semantic gaps between data representations has historically been achieved by arriving at a middleground between rigid standardization on the one hand, and extreme customization on the other. \arch does not change the fundamentals of these tradeoffs and instead merely provides a flexible framework within which we can arrive at this desired  middleground.

\label{subsec:join}
\begin{outline}
\end{outline}

\subsection{Putting the pieces together}
\label{subsec:e2e}
We now step through the ``life of data'' in a \arch implementation of our canonical scenario of a user with a mobile phone interacting with a building app. 

\paragraph{Discovery and ingestion.} Continuing with our notation from earlier, we assume that (for example) the user discovers the building app \Ab by scanning a QR code on entering the building; doing so launches a pop-up via which the user indicates her willingness to share data from \Au with \Ab. This leads to an app-level discovery process in which \Au and \Ab exchange any relevant information (\eg credentials) including the names and API endpoints (\S\ref{sec:impl}) for \Cu and \Cb. Assuming all app-level conditions are met, \Ab invokes \texttt{join(\Cu, \Cb)} on its underlying \arch runtime that implements \Cb. The \arch runtime implements the steps in \S\ref{subsec:join}, as a result of which data from \Cu is ingested and processed by \Cb. How \Cu's data is processed is determined by \Cb's ingress pipeline and the match:action rules relevant to \Cu's schemas. As a specific example, let's say that a data record $R_{user}$ exported by \Cu includes two fields: a unique user ID and a temperature reading in celsius and that \Cb's ingress pipeline both augments and transforms this data into a record $R_{bldg}$ that consists of a building ID, %
user ID, time-stamp, and temperature in fahrenheit. These transformed records are stored in \Cb.store.

\paragraph{Querying contexts.} 
Now consider a campus analyst interested in understanding the usage and temperature conditions in the above building. The analyst issues queries such as {\it "what is the average occupancy and temperature of this building ID between 8am-10am?"}, or {\it "how much time does an average user spend in building ID?"}. Likewise, our user might want to ask {\it what is the warmest floor in the building at times when I visit?"}. Since all data relevant to that building is stored in \Cb.store, these queries can be efficiently executed directly over that single data store, rather than sifting through all user records across all campus buildings in order to identify the relevant data over which to compute statistics. The latter is what would be required in existing systems~\cite{dspace,aws-iot,homeassistant} that simply load input data into per-user device tables.  Moreover, by associating an ingestion pipeline with each building context, \arch makes it easy to augment/modify input data with the required fields needed to optimize query performance \emph{for that context}; \eg the building ID. By contrast, existing systems would need to maintain separate metadata tables mapping a user ID to the building ID they were in at different points in time. In short, \arch's per-context modularity avoids the complexity and inefficiency of having to sift through a sprawl of per-table devices.%

\paragraph{Decentralized control and configuration.} 
Later, the user enters their campus' medical building and, as before, agrees to contribute data to \Ab. However, for HIPAA compliance, this building's operator must store a copy of ingested data records in their insurance provider's cloud. The building operator can easily achieve this by setting up a data replicator function in the ingress pipeline for the medical building's context. Similarly, the user might configure their egress policy to drop or anonymize their user ID when exporting data to this particular building. 
Again, \arch's context-level modularity makes it easy for both users and building operators to unilaterally define and enforce their policies. 
Achieving the same in a system that simply loads all user records into a monolithic campus-wide data lake would be more complicated since the $A_{med}$ operator would need to (manually) communicate their replication needs to the campus operator and the latter would (likely) require complex scripts to first identify which users were in the medical building and at what time and then copy those select records to the cloud.

\paragraph{Developing and configuring a context.}
Finally, let's consider a scenario in which the literature department decides it is time they also offer a building app. The developer tasked with the job might start by simply copying the context implementation from another building app. However, the building operator is unsure what ingestion policies to configure. They can thus start by configuring their match:action table to simply log all input records and analyze the data offline to determine what data types are available and useful. The operator can then use their new understanding to refine their match:action rules for better data curation~\cite{stonebraker2013data} and can continue to do so over the lifetime of the app and as new devices and data types appear. 
In short, \arch allows flexible and \jit ingestion from any available data source, enabling richer insights. Moreover, the context-level modularity simplifies app development since it supports (context) code reuse and simplifies configuring and evolving individual contexts.

\begin{outline}

\end{outline}

\subsection{Other design pieces}
In addition to any app-level security and access control, \arch supports access control at the \emph{context} level.  We achieve this using standard role-based access control techniques for which each context is associated with \emph{role} metadata that gets checked at \join time. Since our RBAC design is fairly standard, we omit further details. 

Beyond \jit data access, we have found that the context abstraction facilitates a number of operational tasks. For example, having a per-context ingress/egress makes it easy to implement policies for tracking provenance, since we can easily log statistics about when and where data was consumed from different sources. Similarly, we can support diverse compliance needs by tailoring how data is stored or replicated for each context; \eg ensuring different standards of encryption and backup for data collected in a public \vs private building. 

Our \arch prototype described in the following section implements RBAC, data provenance, logging, and replication. 

%% file: s4.tex
\section{Implementation}
\label{sec:impl}

\begin{figure}
     \centering
     \footnotesize
     \includegraphics[width = 0.3\textwidth]{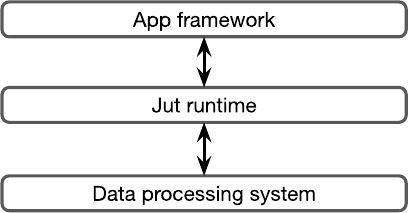}
     \figspace
     \caption{\textbf{Overall system architecture.}}
     \label{fig:cot-arch}
     \figspace
\end{figure}

\arch is implemented as a runtime that sits between an app and its underlying data processing system as shown in Fig.\ref{fig:cot-arch}; together these three components constitute a complete smart-space application or service. %

\subsection{End-to-end System Architecture.}
As shown in Figure~\ref{fig:cot-arch}, \arch integrates with existing systems that implement data processing and application-layer logic. We briefly describe our system choices for these components. 

\paragraph{Zed as the data processing layer.}
\arch uses the recent Zed~\cite{zed-cidr} system as its underlying data processing engine. Zed introduces a new \emph{super-structured} data model that aims to unify the traditional relational and document models in order to achieve the benefits of both -- efficient analytics (the relational model's strength) and flexibility in ingesting and querying heterogeneous data (the document model's strength)?

\arch does not require using Zed; we also considered systems such as Postgres~\cite{postgres}, DuckDB~\cite{duckdb}, and Sqlite~\cite{sqlite}. Ultimately, we selected Zed because of its sophisticated support for heterogeneous data. In Zed, \emph{all} data are strongly typed, as in relational models. %
However, there is no restriction on which types of data may coexist in the same stream of data (akin to JSON systems). Instead, Zed requires that data is ``self-describing'' by storing type definitions inline with the data stream: \ie when a new data type appears in a stream of super-structured data, its type definition is stored inline with the data; this ensures that data consumers always have the necessary type information to parse data.
This approach is convenient for \arch since we can define rich type-based match:action rules; \eg using different actions when encountering a new type definition and processing records differently based on their type. 

\paragraph{dSpace as the application framework.} There are a growing number of frameworks for building IoT and smart-space applications: AWS IoT~\cite{aws-iot}, Samsung SmartThings~\cite{smartthings}, and Comfy~\cite{comfy}. However most are closed systems: we can use their publicly available service APIs but cannot modify their internals. Within open source options, we considered HomeAssistant~\cite{homeassistant}, BOSS~\cite{nsdi13-boss}, and dSpace~\cite{dspace}. We ultimately chose dSpace since it most directly aligns with our focus on smart spaces. In dSpace, developers model a space or sensor using a ``digivice'' abstraction, which is largely equivalent to our notion of a context. A digivice is a \emph{control} abstraction that implements (for example) the automation logic for a room (when and how to dim lights, \etc). For every digivice, dSpace does provide a corresponding ``digidata'' abstraction to model the data associated with a space. In the dSpace implementation, a digidata is simply a wrapper around Postgres with no support for the dynamic ingestion, policies, \etc that we address. We thus modified their digidata abstraction to instead integrate with \arch thus replacing Postgres by our \arch runtime $+$ Zed. An added benefit of this choice is that -- by comparing dSpace-over-Postgres to dSpace-over-\arch -- we can evaluate the benefits of \arch with the same app layer (\S\ref{sec:eval}). 

\paragraph{Legacy data sources.}
In \S\ref{subsec:join}, we described \texttt{join} as having the ingress of a context read data from the output of another. In practice, some contexts will have to read from data sources that do not conform to our architecture and, indeed, we use many such sources in our evaluation in \S\ref{sec:eval}. For such ``legacy'' sources, we rely on proxies that access data from the source based on whatever APIs or protocols they expose - \eg using the Matter protocol~\cite{matter} or via the device vendor cloud~\cite{lifx-light} - and convert the stream to a Zed format~\cite{zed-format}. 
Fortunately, Zed already offers a number of such proxies which we were able to leverage.

\eat{ 
}

\subsection{The \arch Runtime}

\begin{figure}
     \centering
     \footnotesize
     \includegraphics[width = 0.4\textwidth]{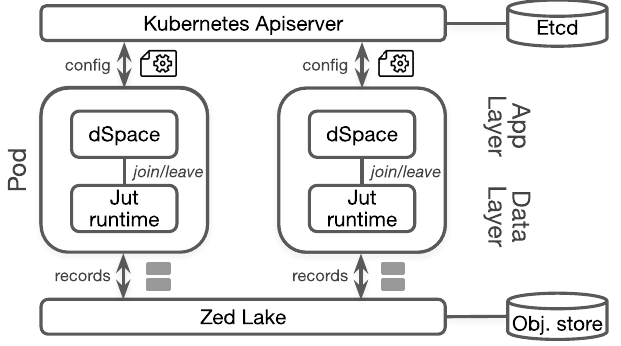}
     \figspace
     \caption{\textbf{\arch's runtime components.}}
     \label{fig:runtime}
     \figspace
\end{figure}

Fig.~\ref{fig:runtime} illustrates the runtime architecture of \arch, comprising the following components:

\pb{Kubernetes.} Kubernetes is a widely used container orchestrator. \arch uses Kubernetes to deploy all of its runtime components and uses its apiserver~\cite{k8s-apiserver}, built on etcd~\cite{etcd}, to store the context metadata, ingress, and egress configurations%
persistently as API objects. \arch also uses the apiserver as the service registry for apps and contexts. 

\pb{Zed lake.} We use the Zed lake~\cite{zed-lake} to support data storage and querying in \arch. The Zed lake stores super-structured data persistently in an object store and organizes data in ``pools''.
\arch leverages the Zed lake by placing each context's data store in a separate data pool; the \arch runtime is responsible for moving data between contexts (pools). 

\pb{App and runtime.} %
The dSpace app and the \arch runtime run in the same Kubernetes pod with one container for each. The \arch runtime exposes the \texttt{join}/\texttt{leave} API via a REST endpoint for the dSpace app to invoke. Meanwhile, the \arch runtime can query and load data from and to the Zed lake whereas apps may query and load data from the \arch runtime via the interfaces exposed by it (\S\ref{sec:prog}). 

We implement a \emph{pipelet data sync agent} in the runtime that continuously pulls data from a data pool on the Zed lake, processes data with the Zed dataflow, and loads it to another data pool on the Zed lake. For each egress-to-ingress pair and data-store-to-egress pair in context, \arch creates an instance of the pipelet agent to plumb the data flow.

Our \arch prototype consists of 2,393 SLOC of Python and 815 SLOC of Go. The \arch programming library, code generators, \texttt{join}/\texttt{leave} extension, and app are written in Python, while the pipelet data sync agent is written in Go. Besides, we developed a Go-based CLI in $\approx$600 SLOC for \arch that includes APIs for context deployment, configuration, and interaction with Zed and dSpace. To implement device connectors for proxied ingestion, we rely on vendor libraries~\cite{ios,emqx,tuyapi}, which account for $\approx$400 SLOC in Python for IoT devices and laptops, and $\approx$550 SLOC in Swift for smartphones~\cite{ios}.

\subsection{Programming Interface}
\label{sec:prog}

\arch exposes a declarative interface built on Kubernetes~\cite{kubernetes}. This interface allows context configurations (metadata, ingress, and egress) %
to be represented as attribute-value pairs, which are stored as API objects on the Kubernetes apiserver~\cite{k8s-apiserver}. We expose context configurations over this interface so that developers and operators can reuse existing Kubernetes tooling to handle configurations. Developers or context operators can specify the ingresses and egresses on the interface. We present the details of the programming interface in the Appendix.

Developers use the join/leave() API in the \arch programming library to inform the context about availability of new data sources. They use the Zed dataflow language~\cite{zed-lang,zed-cidr} to specify the dataflow operators in each context's ingress and egress. Zed provides convenient operators such as filtering and cleaning data using type information (\S\ref{subsec:result}). Users can query the egresses of contexts using the same dataflow language. For example, a building administrator can query the BioHall occupancy by running \texttt{jut query BioHall.egress.occupancy "avg()"} (jut is the CLI). Developers can also use the query API to run a query against the context in the JIT-DA apps; or use the \texttt{@on.context(egress)} (implemented as a Python decorator) to watch and process the data streams continuously, which is useful when implementing data-driven automation. Besides, an app can load data to the $C.store$ by calling \texttt{context.load({..})} with the data records, which can be used for ingesting data from sources that \arch's ingress doesn't already support, such as devices that communicate via custom device driver/libraries.

%% file: s5.tex
\section{Evaluation}
\label{sec:eval}

We evaluate \arch to answer two high level questions: (i) does \arch simplify the use and development of smart-space apps, relative to existing app frameworks? and (ii) what performance overheads does \arch introduce and does our implementation's end-to-end performance meet real-world  requirements?

We explain our experimental setup (\S\ref{subsec:setup}) and then evaluate each of the above questions in \S\ref{subsec:result} and \S\ref{subsec:benchmark} respectively.

\subsection{Experimental setup}
\label{subsec:setup}

Our overall evaluation approach is as follows. Using a set of real-world physical devices, we construct a series of smart-space scenarios, each designed to highlight a particular aspect of \arch's design (\eg querying, implementing policy). We evaluate the complexity of this process using the metrics defined below. 
In each case, we then attempt to implement the same scenario using two existing systems that we select as our baseline and again quantify the complexity of the experience using the same metrics. 
In what follows, we introduce our devices, metrics, and baselines.

\begin{table}
\footnotesize
\centering
\begin{tabular}{|c|c|c|c|}
\hline
\textbf{Device} & \textbf{Brand} & \textbf{Model} & \textbf{Quantity} \\ \hline
Smartphone & Apple & iPhone 13 Pro Max & 2 \\ \hline
Laptop & Apple & Macbook Air M1 & 2 \\ \hline
Motion Sensor & Ring & Alarm Security Kit & 6 \\ \hline
Underdesk PIR & Pressac & \begin{tabular}[c]{@{}c@{}}Wireless desk \\ occupancy detector\end{tabular} & 2 \\ \hline
Contact Sensor & Eve & Door \& Window & 4 \\ \hline
Plug & Eve & Plug \& Power Meter & 4 \\ \hline
\begin{tabular}[c]{@{}c@{}}Hub + Thread \\ Border Router\end{tabular} & Apple & HomePod mini & 1 \\ \hline
Smart light bulb & Lifx & Mini & 8 \\ \hline
Heater + Fan & Dyson & HP01 & 1 \\ \hline

\end{tabular}
\caption{Devices in our testbed.}
\label{tab:device}
\figspace
\end{table}

\pb{Devices.} Our experiments use the physical devices listed in Table~\ref{tab:device}. These devices span personal mobile devices as well as fixed sensors. For the former we use Apple phones and laptops - these devices both contribute data as well as run user-facing apps that consume data from our smart-space contexts.
Two devices -- Pressac~\cite{pressac} and Ring~\cite{ring-motion} -- are used to detect occupancy by signaling whether there is motion or not. 
The Eve Contact sensor~\cite{eve-door} is used to detect door open/close events.
The Eve plugs~\cite{eve-plug} and Dyson~\cite{dyson} heater connect to electrical appliances (\eg TV, fridge, air purifier) and report energy statistics.
Lifx~\cite{lifx-light} is a smart light bulb. 
The table also lists what we term the "native" API via which the vendors export data from their devices. For the iPhone and laptop we wrote a custom \arch client that exports relevant data (described later) in JSON. In our experimental scenarios, these devices are connected via an Apple Hub $+$ Thread border router. 

\pb{Metrics.} 
We evaluate ease of development using source lines of code (\emph{SLOC}). 
We count the source lines of code, configurations, and scripts required to build/configure an app.
To evaluate ease of use, we measure query complexity (\emph{Qcx}) as representing the effort required to write queries. 
Qcx is calculated by summing the number of dataflow operators and data subjects in a query. For example, a query with two data subjects and one dataflow operator, such as \texttt{BioHall.egress.occupancy "avg()"}, has a Qcx of 2.\footnote{We recognize that there is no widely accepted method for measuring query complexity, and hence we adopt this simple approach, inspired by proposals from prior literature~\cite{software-complexity,sql-complexity}.} 

\pb{Baselines.} We compare \arch to two app frameworks: AWS IoT~\cite{aws-iot} and dSpace~\cite{dspace}. Both require an underlying data processing layer. For AWS-IoT, we use AWS Timestream~\cite{aws-timestream} which is a common recommendation and, for dSpace~\cite{dspace} we use Postgres~\cite{postgres} which is the default choice in dSpace. 
We select these as AWS IoT is widely used in industry, while dSpace represents the state-of-the-art from research. In reporting our findings (Table~\ref{tab:cot}), we select the \emph{better} of AWS-IoT and dSpace and report that number, thus providing an optimistic view of baseline performance; our explanations in the text make clear which system is being discussed.

Unless otherwise mentioned, we run the various software components (\arch runtime, Zed lake, dSpace, and the baselines) using Kubernetes (v1.21.0) in a Thinkcentre M720 machine (Intel Core i5-8400T, 6 cores).

\pb{Contexts} 
We developed a \arch context for each of the physical devices and data sources listed in Table~\ref{tab:device}.
These contexts are listed in Table~\ref{tab:setup}\eat{(entries 1-X)},  including the data sources they pull data from and the egresses they expose. 
As expected, the ingress for these contexts is typically the vendor's native API.  In some cases, multiple devices can be represented by the same context. E.g., the Motion context reads data from the Pressac~\cite{pressac} or Ring and internally converts their schemas to the same representation on egress. A benefit of this consolidation is that downstream contexts that consume data from Motion need not address the heterogeneity across the native data sources.  
Finally, we created contexts for higher-level constructs such as room, building, campus, \etc. These contexts consume data from other \arch contexts (\vs native APIs) and, as we'll show, can themselves serve as data sources for downstream contexts.

In counting SLOC, we include all (non runtime) code needed to implement the above contexts but leave out the code specific to converting from a native API format to Zed's format. We omit this code because it is incurred by any solution including our baselines. Moreover, this overhead is somewhat artificial since it depends on the (native and egress) schemas we define,  rather than fundamental differences in framework abstractions. 

\begin{table}[t]
\footnotesize
\centering
\begin{tabular}{|c|c|c|}
\hline
\textbf{Context} & \textbf{Ingress} & \textbf{Egress} \\ \hline
Smartphone & Jut client & \begin{tabular}[c]{@{}c@{}}conn, \\ bssid\end{tabular} \\ \hline
Laptop & Jut client & \begin{tabular}[c]{@{}c@{}}conn, \\ bssid\end{tabular} \\ \hline
Motion & \begin{tabular}[c]{@{}c@{}}Pressac native API \\ Ring native API\end{tabular} & detected \\ \hline
Contact & Matter native API & detected \\ \hline
Lamp & Lifx native API & energy,  brightness \\ \hline
Appliance & \begin{tabular}[c]{@{}c@{}}Matter native API\\ Dyson native API\end{tabular} & energy \\ \hline
Room & \begin{tabular}[c]{@{}c@{}}Motion.detected, \\ Smartphone.conn, \\ *.energy\end{tabular} & \begin{tabular}[c]{@{}c@{}}netSpeed\\ energy\\ occupancy\end{tabular} \\ \hline
Building & \begin{tabular}[c]{@{}c@{}}Room.energy, \\ Room.occupancy\end{tabular} & energy,  occupancy \\ \hline
Campus & \begin{tabular}[c]{@{}c@{}}Building.energy, \\ Building.occupancy\end{tabular} & \begin{tabular}[c]{@{}c@{}}energy, \\ occupancy\end{tabular} \\ \hline
\end{tabular}
\caption{Kinds of contexts used in the experiments.}
\label{tab:setup}
\figspace
\end{table}

\subsection{Implementing smart spaces with \arch}
\label{subsec:result}
 We implemented 7 different smart-space scenarios using the above devices and contexts. 
 Our results showing the SLOC and Qcx for each scenario are summarized in Table~\ref{tab:cot}. Additional detail including code examples can be found in our Appendix.

\begin{table*}[]
\centering
\footnotesize
\begin{tabular}{|c|c|c|c|c|c|}
\hline
\textbf{Scenario:} & \textbf{S1} & \textbf{S2} & \textbf{S3} & \textbf{S4} & \textbf{S5} \\ \hline
\textbf{Requirement} & \begin{tabular}[c]{@{}c@{}}Query-over-\\ context\end{tabular} & \begin{tabular}[c]{@{}c@{}}Context-as-\\ a-source\end{tabular} & \begin{tabular}[c]{@{}c@{}}Opportunistic\\ Ingestion\end{tabular} & \begin{tabular}[c]{@{}c@{}}Data-driven\\ automation\end{tabular} & \begin{tabular}[c]{@{}c@{}}Handling data \\ heterogeneity\end{tabular} \\ \hline
\textbf{\begin{tabular}[c]{@{}c@{}}SLOC\\ Jut (baseline)\end{tabular}} & 25 (-) & 19 (-) & 12 (178) & 14 (45) & 8 (-) \\ \hline
\textbf{\begin{tabular}[c]{@{}c@{}}Qcx\\ Jut (baseline)\end{tabular}} & 18 (159) & 6 (73) & 6 (43) & 11 (33) & 4 (-) \\ \hline
\end{tabular}
\caption{\textbf{Comparison of the ease-of-development and ease-of-use between \arch and baselines.} Baseline results are shown in parentheses. \textbf{SLOC:} total lines of source code and configurations required to implement components for each scenario. \textbf{Qcx:} total query complexity for completing analytics tasks in each scenario.}
\label{tab:cot}
\figspace
\end{table*}

\noindent \underline{\bf S1: Query by context} 
This scenario focuses on the ease of supporting query by context. Our goal is to query rooms for their energy usage and occupancy, identify the rooms that are least occupied or consume the most energy, and analyze the correlation between energy use and occupancy. 

To set up this experiment, we installed devices in 4 rooms/offices, where each room has one contact sensor, two motion detectors, two lamps, and one appliance; Each room has its own WiFi access point (AP). We reuse this room setup in subsequent scenarios.

We then ran three types of queries over the room contexts: (i) querying an individual room (for its current occupancy or energy usage), (ii) querying across multiple rooms (to rank rooms by their occupancy or energy usage), and (iii) querying across a room's different egresses (to identify the correlation between a room's energy usage and occupancy). For example, for (ii), we rank and find the least used room with:
\mintinline{shell}{> jut query room1.egress.occupancy "head | sort occupancy"}

Implementing this scenario required 25 SLOC, most of which were to implement the necessary contexts. Composing data sources required few SLOC - calls to \join and, as necessary, configuring context match:action rules. 

In terms of query complexity: implementing the above queries with \arch required a total of 18 Qcx. 
By contrast, using dSpace with Postgres, we first need to manually create a table that stores the mapping between devices and their room. Given this, we can then run this query but it is still inefficient in Qcx. The same queries required 159 Qcx because dSpace stores input data in per-device Postgres tables. The user thus first has to filter table rows that belong to the room of interest and additionally aggregate those rows over time to derive the room occupancy values (detailed queries are in our Appendix). By contrast, in \arch, the relevant data is already selected and stored in the room context's store which can be directly queried.

\noindent \underline{\bf S2: Context as a data source} 
Now consider that we want to enable building-level energy and occupancy analysis. A convenient approach is to compose the room contexts to a building context that computes and exposes building-level data from the room-level data.
Implementing this in \arch took an \emph{additional 19} SLOC (over S1) to implement the building context: 
we specify the room's \texttt{occupancy} and \texttt{energy} egresses as the building's ingresses and the building's ingress dataflow aggregate input data as it arrives. 
Querying for building occupancy or energy are then straightforward queries to \texttt{room1.egress.occupancy} or \texttt{room1.egress.energy}, respectively, with a Qcx of 6.
Implementing this query with Postgres incurs a higher Qcx for similar reasons as above.

\noindent \underline{\bf S3: Ingestion from \jit sources} 
This scenario focuses on \arch's ability to opportunistically ingest data from user devices. Here, our smart campus app allows people to identify rooms with both good network connectivity. To achieve this, it consumes network quality data from tenants' phones and laptops. 
We now extend the scenario from above to include phones and laptops as potential data sources for a room. 

Our user phone app is configured to \join and \leave rooms upon detecting a change in the BSSID for the WiFi AP it is currently connected to. Each time the phone changes APs, it takes a network measurement in the background using the NDT tool~\cite{ndt,ndt-ios} and reports the raw data to the phone's context 
(exposed by \texttt{phone.egress.netTest}). We add an ingress to the room context to ingest data from \texttt{phone.egress.netTest}, and an egress \texttt{room.egress.netSpeed} that exports the average (across all phones/laptops) download/upload bandwidth data to the building context, and so on. We repeat the same process for the laptop and its context.

Implementing the above scenario in \arch was straightforward, because much of the low-level "plumbing" code is encapsulated within the \join API.
Thus supporting this scenario required creating a context for the phone/laptop with a \texttt{phone.egress.netTest} egress; this took 3 SLOC each. Similarly, adding the \texttt{room.egress.netSpeed} egress to the room context required 5 SLOC and 1 match:action rule to room's ingress. 
To find rooms with good connectivity, users simply query the room for their average connectivity quality, which is achieved with a Qcx of 6. 

By contrast, this scenario cannot be easily supported in AWS-IoT or dSpace.
In order to approximate \jitda in AWS-IoT we hack together a solution as follows. 
AWS-IoT allows devices (together with their schemas and other metadata information) to be registered in a device registry. We create an entry for our phone and laptop devices and manually hardcode an associated room identifier (RoomId) for each device (deviceID). We then write a script that, when it receives an app-level signal that a device is in the room, takes the following steps: (i) provision a new table in AWS TimeStream for that device, (ii) updates the AWS message broker with a new topic corresponding to the deviceID in the device's data stream such that all records from the device are routed to the new table, (iii) set up a rule associated with the topic that allows incoming records to be processed prior to storage in TimeStream; this rule adds a new field to each data record into which we write the RoomId corresponding to deviceID,  retrieved from the device registry. (Note that supporting user mobility would require additional techniques to dynamically update the deviceID-to-roomID mapping in the device registry.) 
Implementing the above took 178 SLOC. 
A query to compute the average connectivity quality of the room then requires averaging the appropriate rows from each device's table, for all devices that might have been in the room. Writing such a query involved a Qcx of 43.

\noindent \underline{\bf S4: Improved app-data integration.} 
We focus now on the ease with which \emph{applications} can 
leverage insights from \jit data: \ie for \emph{data-driven} automation.
For this scenario, our goal is to support two tasks. First, we want a room's brightness level to be automatically set when the door opens, where the level of brightness is computed from the historical data for that room. 
Second, as a security measure, we want an app that automatically sends an SMS to a home owner when room occupancy exceeds a specified threshold. 
In \arch, the above automation rules are specific in dSpace but rely on queries made to the underlying \arch data processing system. Writing these queries in \arch is straightforward because the abstraction over which dSpace expresses the above automation logic -- \ie a room -- is clearly mirrored in the underlying data storage. Hence, we simply write dSpace's automation logic to make continuous queries to the relevant room context - this took 14 SLOC (in dSpace) and 11 Qcx to query the room brightness and occupancy. By contrast, implementing the same over Postgres took 3x more Qcx and 3.2x more SLOC for reasons similar to those discussed above (provisioning and querying over multiple per-device tables).

\noindent \underline{\bf S5: Handling heterogeneous data.} 
Our last scenario highlights the issue of heterogeneous data sources. 
Our goal here is a smart-app that wants to allow users to track their personal carbon footprint: \ie allowing users to track the energy consumption of (say) lamps, heaters, \etc that they encounter at different locations on campus during their stay. To achieve this, the app must be able to opportunistically consume data from a heterogeneous set of (device, space, building )contexts as they move. 

To implement this scenario, we configure the match:action rules on the phone context to ingest and aggregate readings from any available data source with a \texttt{*.egress.energy}
These data may have heterogeneous and unexpected/unknown schemas, such as data types (\eg string vs. float), names (energy'' vs. power''), or units (\eg watt'' vs. kilowatts''). For example:

\begin{minipage}{.5\textwidth}
\minttopspace
\begin{minted}[xleftmargin=0.075in,fontsize=\footnotesize]
{shell}
{watt:"80",from:"biolab",event_ts:..,ts:..}
{watt:100,from:"office",event_ts:..,ts:..}
{power:120.,unit:"watt",from:"lounge",event_ts:..,ts:..}
\end{minted}
\end{minipage}

Contexts in \arch are able to ingest these heterogeneous data records. At the phone context's ingress, we clean and convert the energy readings to the same data type and unit with:
\begin{minipage}{.5\textwidth}
\minttopspace
\begin{minted}[fontsize=\footnotesize]
{yaml}
flow: "rename watt:=power | shape(this, <{watt:float64}>) 
      | cut watt,event_ts,from"
\end{minted}
\end{minipage}

Querying for a user's carbon footprint is then a trivial query to the energy egress of the user's phone context. We implemented this in \arch with 8 SLOC and a Qcx of 4. 

In summary, by introducing the modularity of contexts and by systematizing how new sources are added/removed to data pipelines, \arch achieves a significant reduction in both development effort (3.2-14.8x SLOC) and query complexity (3-12x Qcx) relative to AWS-IoT and dSpace.

\subsection{Performance benchmarks}
\label{subsec:benchmark}

We aim to answer three performance-related questions for \arch: (1) How does \arch's context-oriented approach impact query performance compared to device-oriented approaches? (2) What is the overhead of using \arch for data-driven insights in the existing applications? and (3) Can \arch scale to a large number of contexts with our implementation choices? 

\begin{figure}
     \centering
     \footnotesize
     \includegraphics[width = 0.39\textwidth]{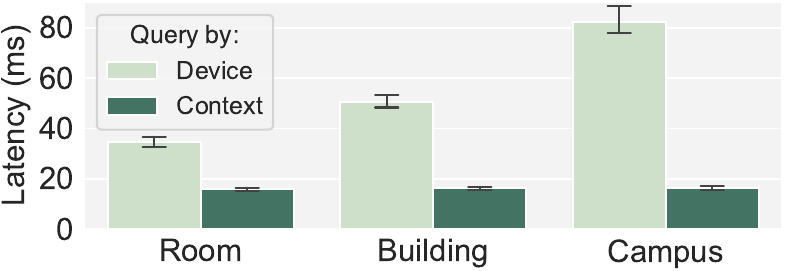}
     \figspace
     \caption{Comparison of query latency when querying over device data (Device) vs. over context data (Context).}
     \label{fig:query}
     \figspace
     \figspace
\end{figure}

\pb{Impact on query performance.} To answer (1), we set up a benchmark with the smart campus app, using a week-long occupancy dataset from the UCI repository~\cite{occupancy-dataset}. We load the dataset into the motion detector contexts, which in turn populate the contexts of rooms, buildings, and the campus. We compare the latency (10 runs, reported average in $ms$) of querying the room, building, and campus occupancy using \arch and Zed lake as the baseline, as \arch already uses it as the underlying analytics engine. For \arch, we wait for the occupancy detector contexts to propagate to the other contexts, whereas for the baseline, we pre-process the dataset to add context information and then load it into Zed lake.
 
Fig.\ref{fig:query} shows that \arch achieved an average latency of $15.8 ms$ for the room context, which is 2.2x faster than the device-oriented baseline. As the context level goes higher, the gap widens with 3.1x for the building context ($16.2 ms$ vs. $50.4 ms$) and 5.1x for the campus context ($16.2 ms$ vs. $50.4 ms$). The performance improvements stem from two factors. First, a \arch context contains only data ingested for that context, so querying the context directly involves loading less data compared to querying all device data, resulting in a 23\% difference in data loading time. Second, querying the context directly enables simpler queries to be evaluated compared to querying device-level data directly. Note that this improvement does not mean that \arch eliminates the need to process data, but instead shifts the processing to contexts, where the data is transformed \emph{on write/ingestion} to enable faster query evaluation.

\begin{table}[]
\centering
\footnotesize
\begin{tabular}{|c|c|c|c|c|}
\hline
\textbf{Setup} & \textbf{Lamp} & \textbf{LR} & \textbf{OLR} & \textbf{OR-SMS} \\ \hline
\textbf{dSpace} & 188.84 ms & 406.91 ms & 736.01 ms & 3.02 s \\ \hline
\textbf{\arch} & 199.01 ms & 419.39 ms & 742.32 ms & 2.88 s \\ \hline
\textbf{Overhead} & 5.40 \% & 3.10\% & 0.90\% & -4.6\% \\ \hline
\end{tabular}
\caption{Overhead of integrating \arch with control applications in the existing IoT framework (dSpace). {\bf Lamp:} Automating lamp brightness. {\bf LR:} Automating room brightness. {\bf OLR:} Room brightness based on occupancy. {\bf OR-SMS:} Overcrowding alert.}
\label{tab:overhead}
\figspace
\end{table}

\pb{Integrating with existing control apps.} We compare the performance of \arch with control applications written in dSpace, which do not have a data-driven component. We reused the setup from scenario S4 (smart home, \S\ref{subsec:result}) to implement three scenarios in \arch and in dSpace. Table \ref{tab:overhead} shows that \arch adds at most 5.6\% latency overhead across the scenarios for the extra latency ($\approx 10 ms$) performing a query against the context. This querying overhead is negligible for control applications whose routines typically take several seconds, such as invoking the Twilio API to send an SMS message upon an overcrowding event. Therefore, the performance impact of \arch on control applications is insignificant.

\begin{figure}
     \centering
     \footnotesize
     \includegraphics[width = 0.37\textwidth]{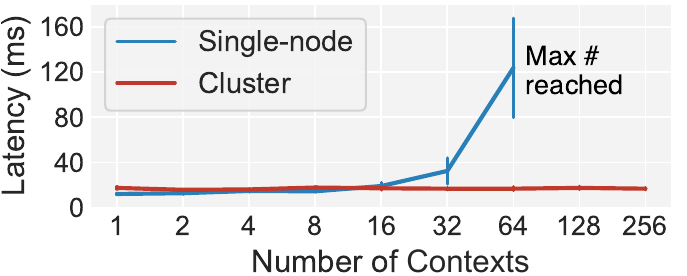}
     \figspace
     \caption{\textbf{Scaling \arch to more contexts.} Comparing the query performance (avg. latency) with increasing number of contexts when deployed in a single-node vs. a cluster with auto-scaling enabled.}
     \label{fig:scale}
     \figspace
\end{figure}

\pb{Scale to large deployment.} \arch's target apps can have various scales from smart homes with a few rooms to buildings and campuses with hundreds of rooms. We assess whether \arch is able to handle \emph{any} of these scales. To that end, we compare the query latency with the single-node setup and a cluster setup. For the latter, we run \arch on a Kubernetes cluster running on AWS EC2 and configure a cluster autoscaler~\cite{k8s-cluster-autoscaler} to automatically provision new EC2 instances (\texttt{m5.xlarge}) under load. We run the smart campus application in S1 and S2 but now vary the number of rooms from 1 to 256; we simulate the data exchange between contexts by feeding the occupancy data to the motion sensor contexts as we did previously. As shown in Fig.\ref{fig:scale}, the results suggest that while the single-node setup can't scale beyond $\approx$60 rooms (due to the per-node resource limits on number of containers), \arch is able to scale to a much larger deployment size (256 rooms on 5 EC2 instances) with a cluster setup. 

%% file: related.tex
\section{Conclusion and Related Work}
\label{sec:related}
This paper presents \arch, a new application framework that simplifies the use and development of apps that perform JIT-DA, with a first-class abstraction for context and explicit app-data coordination. Besides the related work discussed in \S\ref{sec:s2} and \S\ref{sec:eval}, we highlight additional works on IoT and database systems that are pertinent to JIT-DA apps.

\pb{Smart space and IoT frameworks}. Today's IoT frameworks provide extensive support for the IoT app layer functionalities, such as tracking component metadata and relationships and implementing actuation logic. For example, AWS IoT Things Graph provides an entity abstraction~\cite{aws-iot-things-graph}, Home Assistant provides a component abstraction~\cite{homeassistant}, and dSpace~\cite{dspace} introduces a ``digivice'' abstraction. However, these frameworks offer only limited support for the data layer of an app. Most frameworks use standalone off-the-shelf data processing systems to store the data streams generated by individual devices, and thus context-related insights must be extracted from these per-device streams. For example, dSpace on Postgres~\cite{postgres} and TimescaleDB~\cite{timescale} (via a ``digidata'' thin wrapper~\cite{dspace}), Home Assistant relies on SQLite~\cite{sqlite}, AWS IoT on Timestream~\cite{aws-timestream}, DynamoDB~\cite{dynamodb}, or IoT Analytics cloud services~\cite{aws-iot-analytics}. These frameworks also lack the necessary abstractions to adapt data processing pipelines to handle the JIT-DA data sources in smart space applications. This results in constrained data sources an app can incorporate or the need for a significant amount of manual configuration and custom code. \arch addresses these issues by providing flexible and adaptable data layer abstractions to support JIT-DA.

\pb{Databases and dataflow systems}. There is extensive work on dataflow systems, including continuous queries~\cite{telegraphcq}, incremental dataflow~\cite{differential}, dataflow for ML workloads~\cite{osdi16-tensorflow}, streaming engines~\cite{spark-stream,flink}, databases~\cite{materialize,risingwave,noria}, and general-purpose analytics engines~\cite{naiad,spark,apache-beam,flink}. \arch, as a new data architecture, and context, as a new data processing abstraction, are complementary to these pioneering approaches. One can leverage these existing techniques in context to support additional dataflow features such as streaming~\cite{apache-beam}, more comprehensive timeseries analytics~\cite{timescale}, and optimizations that improve context performance and efficiency by supporting stateful operators and incremental computation~\cite{differential,noria,telegraphcq}. In turn, \arch can help scale these systems to handle JIT-DA and support diverse apps. We plan to explore integration of these systems and techniques with \arch in future work.

%% file: supplement.tex
\begin{appendices}
\section*{Appendix}
\label{appendix}

\subsection{Programming Interface for \arch}
\label{apx:prog}

\begin{table}[h]
\footnotesize
\centering
\begin{tabular}{|l|c|c|}
\hline
\multicolumn{1}{|c|}{\textbf{API}} & \textbf{At} & \textbf{Description} \\ \hline
c.query(query, egress="main") & App & Query an egress. \\ \hline
c.load({"spl":.., "unit":..}) & App & Load data to store. \\ \hline
on.ctx(egress="main") & App & Watch an egress. \\ \hline
sources: [ kind | name.egress, .. ] & Context & Specify data sources. \\ \hline
flow: "head | cast(spl,<string>)" & Context & Declare dataflow. \\ \hline
jut query c.egress [ .. ] "sort watt" & CLI & Query egress(es). \\ \hline
\end{tabular}
\caption{\textbf{Key APIs in \arch and their usage.} context: Context data router; app: Context controller; CLI: \arch command line.}
\label{tab:api}
\end{table}

We provide an overview of the programming and analytics APIs in \arch, using (simplified) examples from a smart campus app in our evaluation \S\ref{sec:eval}. Table~\ref{tab:api} summarizes the programming interfaces available in \arch.

\begin{figure}[h]
\footnotesize
\begin{minted}{yaml}
metadata:
  kind: jut/v1/Building
  name: BioHall
ingress:
  room_energy:
    intent: [ room.energy ]
    sources: [ BioLab.energy, Lounge.energy ]
    flow: "room_watt := sum(watt)"
  room_occupancy:
    intent: [ room.occupancy ]
    sources: [ BioLab.occupancy, Lounge.occupancy ]
    flow: "sort -r event_ts | head"
    flow_agg: |-
      put total:=count() |
      occupancy > 0 |
      room_occupancy := cast(count(),<float64>)/total
    patch_source: true # add field "from: source"
egress:
  energy:
    flow: "watt := room_watt"
  occupancy:
    flow: |-
      room_occupancy | 
      rename room_occupancy := occupancy
\end{minted}
\caption{\textbf{Example context configuration} for a building context that generates the energy and occupancy data of BioHall.}
\label{code:building}
\figspace
\end{figure}

\pb{Programming context.} \arch's context exposes a declarative interface built on Kubernetes~\cite{kubernetes}. This interface allows configurations to be represented as attribute-value pairs, which are stored as API objects on the Kubernetes apiserver~\cite{k8s-apiserver}. We extend this interface to support context configurations, so that developers and operators can reuse existing Kubernetes tooling to handle configurations. Fig.\ref{code:building} shows the programming interface for the building BioHall. Developers or context operators can specify the ingresses and egresses on the interface, including the name, dataflow, and data sources (for ingress). Specifically, lines 4 to 16 contain the configurations for two ingresses, \texttt{room\_energy} and \texttt{room\_occupancy}, where the BioHall ingests room-level energy and occupancy as specified in the \texttt{intent} fields (lines 6 and 10) with indirect references and are resolved in the \texttt{sources} (lines 7 and 11). \arch also allows the use of an ``any'' quantifier for context, as shown in Fig.\ref{code:phone-lamp} (right; line 6), which allows AlicePhone to collect energy readings from any contexts it may join. 

\begin{figure}
\centering
\footnotesize
\begin{minipage}{.21\textwidth}
\begin{minted}{yaml}
metadata:
  kind: jut/v1/Lamp
  name: CeilingLamp 
# control config
control:
  power:
    intent: "on"
  brightness:
    intent: 0.2
# data config
egress:
  energy:
    flow: |-
      watt | 
      cut watt,event_ts
\end{minted}
\end{minipage}
\begin{minipage}{.25\textwidth}
\begin{minted}{yaml}
metadata:
  kind: jut/v1/Phone
  name: AlicePhone
ingress:
  footprint:
    intent: [ *.energy ]
    sources: [ BioLab.energy ]
    flow: "sort -r event_ts | head"
    flow_agg: "watt := sum(watt)"
    eoio: false
egress:
  footprint:
    flow: "avg(watt) by every(1h)"
  noise: 
    flow: "spl | cut spl,event_ts"
\end{minted}
\end{minipage}
\caption{\textbf{Example context configurations.} Left: A lamp context. Right: A phone context.}
\label{code:phone-lamp}
\figspace
\end{figure}

\begin{figure}
\footnotesize
\begin{minted}[xleftmargin=0.0in,linenos=false,]{bash}
{room_energy:80,unit:"watt",event_ts:..,ts:..}
{room_energy:120,unit:"watt",event_ts:..,ts:..}
{room_occupancy:0.5,event_ts:..,ts:..,from:"biolab1,lounge"}
{room_occupancy:1.0,event_ts:..,ts:..,from:"biolab,lounge"}
\end{minted}
\vspace{-0.1in}
\caption{Example data from building BioHall's data store, with timestamps (``event\_ts'' and ``ts'') omitted for brevity.}
\label{fig:building-records}
\figspace
\end{figure}

\pb{Processing context data.} Fig.\ref{fig:building-records} shows example data records from the BioHall context with the Zed data format. These records have different schemas and are ordered by their processing time \texttt{ts}, with the values omitted for brevity. In \arch, each data record contains two timestamps - the event timestamp (\texttt{event\_ts}) and the processing timestamp (\texttt{ts}). The event timestamp tracks when the record is generated at the data source. If the data source doesn't attach an event timestamp to the record, then the first-hop context will set the \texttt{event\_ts} to be the \texttt{ts}. The \texttt{ts} on the other hand tracks when the context processes the data, \ie when the ingress loads it to the store. 

Apart from being used in analytics such as timeseries queries, the timestamps are also used to implement exactly-once guarantees. The pipelet agent supports exactly-once, in-order processing (EOIO) when pulling data records from one Zed lake pool/branch to another. This is achieved by writing data records along with the latest \texttt{ts} in the records as part of the commit message atomically, which is then stored at the target pool. Upon failure and restart, the agent will first read and cache the last-seen \texttt{ts} from the target pool's commit messages. When querying data from the source pool, the agent filters the data by reading only the ones with \texttt{ts} greater than the latest \texttt{ts}. 

\subsection{Queries in Performance Benchmarks}
\label{apx:queries}

\begin{figure}[h]
\footnotesize
\inputminted{shell}{./code/queries.sh}
\caption{Queries for room, building, and campus occupancy on the device-level data.}
\figspace
\label{code:queries}
\end{figure}

\subsection{Room context}
\label{apx:room}
\begin{figure}
\footnotesize
\begin{minted}{yaml}
kind: Room
metadata:
  name: BioLab
meta:
  auto_brightness: false
ingress:
  device_energy:
    sources: [ *.energy ]
    flow: "sort -r event_ts | head"
    flow_agg: "select sum(watt) as watt"
    eoio: false
  noise:
    sources: [ phone.spl ]
    flow: "sort -r event_ts | head"
    flow_agg: "select avg(spl) as noise"
    eoio: false
egress:
  energy:
    flow: "watt | cut watt,event_ts"
  occupancy:
    flow: "cut occupancy,event_ts"
  people_count:
    flow: |- 
      num_human | cut num_human,event_ts 
      | rename count:=num_human
noise:
  flow: "noise | cut noise,event_ts"
\end{minted}
\caption{Configurations of a room context data router that generates building energy and occupancy data based on occupancy sensors and devices in the room.}
\label{code:room}
\end{figure}

Developers use the Zed dataflow language~\cite{zed-lang,zed-cidr} to specify the dataflow in ingresses and egresses. Zed provides convenient dataflow operators such as filtering and cleaning data using type information. Note that each ingress has two attributes for dataflow specification (lines 10 to 14). The dataflow in the \texttt{flow} attribute processes data records from each data source, while the one in the \texttt{flow\_agg} attribute processes all records combined from all upstream paths (\ie the resultant data records from \texttt{flow}). Egresses are specified similarly (\eg line 19 that renames the occupancy data), except that an egress does not have \texttt{flow\_agg} nor \texttt{sources} since it reads from the context's data store only.

\pb{Querying over contexts.} Users can query the egresses of contexts using the same dataflow language used for programming the contexts. For example, a building administrator can query the BioHall occupancy by running \texttt{\arch query BioHall.occupancy "avg(occupancy)"}. Developers can also use the query API in app to run a query against the context; or use the \texttt{.on.context(egress)} (implemented as a Python decorator) to watch and process the data streams continuously, which is useful when implementing data-driven automation. Besides, app can load data to the $R.store$ by calling \texttt{context.load({..})} with the data records, which can be used for ingesting data from sources that \arch's proxied ingestion doesn't already support, such as devices that communicate via custom device driver/libraries. For example, in Fig.\ref{code:phone-lamp}, AlicePhone may collect noise data (measured in sound-pressure-level, \emph{spl}), which are first sent to the app and then loaded into the context. More examples of analytics using \arch are presented in \S\ref{sec:eval}. 

\subsection{Query-by-context over Device Data}
\label{apx:query-device}

\begin{figure}
\footnotesize
\begin{minted}{sql}
CREATE VIEW temp
AS
  SELECT Max(time) AS max_time,
         room,
         id
  FROM   device_table
  WHERE  measure_name = 'Underdesk'
  GROUP  BY room,
            id;

CREATE VIEW most_recent
AS
  (SELECT device_table.room,
          device_table.measure_name,
          device_table.detected,
          device_table.id
   FROM   device_table
          JOIN temp
            ON temp.max_time = device_table.time
               AND temp.room = device_table.room
               AND temp.id = device_table.id
   GROUP  BY device_table.room,
             device_table.measure_name,
             device_table.detected,
             device_table.id);

SELECT Count(*),
       room AS occupancy
FROM   most_recent
WHERE  detected = true
GROUP  BY room 
\end{minted}
\caption{Querying room occupancy over device data in Postgres.}
\label{code:apx-room}
\figspace
\end{figure}

\begin{figure}
\footnotesize
\figspace
\figspace
\begin{minted}{sql}
CREATE VIEW temp
AS
  SELECT Max(time) AS max_time,
         room,
         id
  FROM   device_table
  WHERE  measure_name = 'Underdesk'
  GROUP  BY room,
            id;

CREATE VIEW most_recent
AS
  (SELECT device_table.room,
          device_table.measure_name,
          device_table.detected,
          device_table.id
   FROM   device_table
          JOIN temp
            ON temp.max_time = device_table.time
               AND temp.room = device_table.room
               AND temp.id = device_table.id
   GROUP  BY device_table.room,
             device_table.measure_name,
             device_table.detected,
             device_table.id);

CREATE VIEW occupancy
AS
  SELECT Count(*),
         room
  FROM   most_recent
  WHERE  detected = true
  GROUP  BY room;

CREATE VIEW total_amt
AS
  SELECT Count(*),
         room
  FROM   most_recent
  GROUP  BY room;

CREATE VIEW total_num_rooms
AS
  SELECT Count(DISTINCT room)
  FROM   device_table;

SELECT Count(DISTINCT total_amt.room) 
    / Sum(total_num_rooms.count)
FROM   occupancy,
       total_num_rooms,
       total_amt
WHERE  occupancy.room = total_amt.room
       AND occupancy.count / total_amt.count > 0.2 
\end{minted}
\caption{Querying building occupancy over device data.}
\label{code:apx-building}
\end{figure}

\end{appendices}